\documentclass{ifacconf}

\usepackage{graphicx}      
\usepackage{natbib}        
\usepackage{amsmath, amssymb, amsfonts} 
\usepackage{xcolor} 
\usepackage{comment} 

\definecolor{matlabblue}{rgb}{0, 0.4470, 0.7410}
\definecolor{matlabred}{rgb}{0.8500, 0.3250, 0.0980}
\definecolor{matlabgray}{rgb}{0.5, 0.5, 0.5}
\definecolor{matlablg}{rgb}{0.75, 0.75, 0.75}
\definecolor{matlabyellow}{rgb}{0.9290, 0.6940, 0.1250}
\definecolor{matlabpurple}{rgb}{0.4940, 0.1840, 0.5560}
\definecolor{matlab_br_red}{rgb}{1, 0, 0}
\definecolor{matlab_br_blue}{rgb}{0, 0, 1}
\definecolor{matlabblack}{rgb}{0, 0, 0}

\begin{document}
\begin{frontmatter}

\title{Learning of LPV Models With Linear Fractional Representations: Initialization, Optimization, and Conditioning \thanksref{footnoteinfo}} 

\thanks[footnoteinfo]{This work is funded by the European Union (ERC, COMPLETE, 101075836), by the Air Force Office of Scientific Research under award number FA8655-23-1-7061, and the MathWorks Inc. Views and opinions expressed are however those of the author(s) only and do not necessarily reflect those of the European Union or the European Research Council Executive Agency. Neither the European Union nor the granting authority can be held responsible for them.}
\thanks[corauth]{Corresponding author.}
\author[First]{Roel Drenth\thanksref{corauth}}
\author[First]{Maarten Schoukens} 
\author[First,Second]{Roland T\'{o}th}

\address[First]{Control Systems Group, Dept. of Electrical Eng., Eindhoven University of Technology, Eindhoven, The Netherlands.(e-mail: \{r.drenth, m.schoukens, r.toth\}@tue.nl).}
\address[Second]{Systems and Control Laboratory, HUN-REN Institute for Computer Science and Control, Budapest, Hungary}
\begin{abstract} 
Linear Parameter-Varying (LPV) models with Linear Fractional Representation (LFR) can describe a wide class of nonlinear systems with a control-friendly structure. However, identifying these models together with their scheduling map from input-output data requires solving a non-convex optimization problem that is highly sensitive to initialization. Consequently, existing methods often rely on many random initializations. Instead, our work proposes an initialization strategy based on the Best Linear Approximator (BLA). Furthermore, motivated by an analysis of the impact of rational-dependency LPV-LFR models on the conditioning of the non-convex cost, two multiple shooting methods that improve the conditioning, unconstrained multiple shooting (UMS) and encoder-based multiple shooting (EBMS), are compared. The initialization and multiple shooting methods are evaluated on a nonlinear mass-spring-damper example, showing that BLA initialization consistently yields accurate results, whereas random initialization has a large spread. The results also show that both UMS and EBMS reduce the spread for randomly initialized models. UMS is simple to tune and well suited to few sub-records, whereas EBMS can exploit many short sub-records to further improve conditioning, at the cost of training an encoder and tuning additional hyperparameters.
\end{abstract}

\end{frontmatter}

\section{Introduction}
As demands on performance of controlled systems in engineering increase, using typical \emph{Linear Time-Invariant} (LTI) assumptions to model and control systems which are often inherently \emph{nonlinear} (NL) limits achievable  performance.

Consequently, there is a need for a framework which is capable of capturing the increasingly important nonlinear nature of systems, yet supports control methods for which performance guarantees can be given. One solution is the use of the \emph{Linear Parameter-Varying} (LPV) framework. LPV models are linear in the input, with a model that varies according to a \emph{scheduling variable}. Through change in the scheduling variable, time-varying and nonlinear behaviour can be captured. However, by virtue of its partially linear structure, controller synthesis with performance guarantees is made possible (see, e.g., \cite{mohammadpour2012}).

Data-driven modelling, i.e., \emph{system identification}, of LPV models is still an area of active research. Traditional LPV identification methods, where access to measured scheduling signals is required, are well developed (see e.g., \cite{toth2010, vanwingerden2009}). However, modern approaches, which identify nonlinear systems by joint estimation of LPV models and scheduling maps, through \emph{Prediction Error Minimization} (PEM), are still receiving a lot of attention. In particular \emph{deep-learning}-based identification of \emph{self-scheduled} methods (e.g. \cite{verhoek2022, bemporad2025b, drenth2026}) have been shown to be capable of accurately modelling nonlinear systems.

Even though these methods are capable of modelling nonlinear systems with high accuracy, the reliability of these identification methods is unclear. In particular, \cite{bemporad2025b} and \cite{drenth2026} rely on selecting best-performing models from a range of randomly initialized models, indicating that the strongly non-convex optimization problems are sensitive to the chosen parameter initialization. In particular, the additional complexity of rational LPV-LFR models worsens the conditioning of the optimization problem.

Despite the importance of initialization in non-convex optimization, this subject has not received much attention in LPV identification literature. Methods using \emph{subspace identification} exist in the \emph{externally-scheduled} case \citep{lee1999, verdult2003}, but require measured scheduling signals, and hence are incompatible with the self-scheduled methods considered in this work. However, use of the \emph{Best Linear Approximator} (BLA) \citep{pintelon2012} has been shown to be effective for a wide range of  nonlinear model structures, including nonlinear input-output \citep{sjoberg1997}, block-oriented \citep{ schoukens2017}, nonlinear state-space \citep{paduart2010, schoukens2021}, and nonlinear LFR models \citep{vanbeylen2013, schoukens2020}. Moreover, \cite{schoukens2021} shows that BLA initializations reduce variance of the obtained accuracy over random initialization methods. Nonetheless, BLA initialization methods have not been explored for LPV model structures.

While initialization methods can improve attainable results in identification, they do not address the ill-conditioning of the cost functions. Moreover, a sufficient BLA estimate cannot always be attained. Consequently,  \emph{multiple shooting} \citep{bock1983} is explored as a second method to improve the reliability of identification.

Simulation-error methods require evaluation of the prediction error across the entire trajectory of the training data. However, \cite{ribeiro2020} has shown that the smoothness (conditioning) of the cost function depends on the recursion depth (simulation length) of the data, especially for models that are locally non-contractive. Multiple shooting aims to limit the recursion depth, by evaluating multiple shorter sub-records, improving smoothness of the cost. Additionally, evaluation of multiple sub-records can be performed in parallel, which leverages modern computing hardware more efficiently.

The use of multiple shooting comes with a trade-off: simulating recurrent models requires knowledge (or estimates) of the initial state for each sub-record. To resolve this, a simple solution is to incorporate the additional initial states into the optimization problem. To ensure continuity between sub-records, constraints can be used in conjunction with constrained optimization solvers \citep{bock1983, ribeiro2017}. Alternatively soft-constraints \citep{forgione2021} can be employed. However, the use of (soft-)constraints reduces computational benefits of MS as gradient computations still require back-propagation along the full trajectory. Instead, using unconstrained multiple shooting \citep{decuyper2020} has been shown to be effective as well. Some methods also opt to use an encoder to produce the initial states from IO data \citep{beintema2023}.

By introducing a separate \emph{encoder network}, \emph{Encoder-Based Multiple Shooting} (EBMS) methods are capable of initial state estimation for any number of sub-records, at a fixed parameter cost. This enables the use of overlapping sub-records, improving \emph{data-efficiency}. Moreover, the large number of possible sub-records that can be used also facilitates the use of mini-batching methods. These features have led to EBMS being used effectively on a wide range of model structures, including LPV state-space models \citep{verhoek2022}. However, EBMS methods come with a few downsides; a separate encoder network needs to be trained, often requiring a large number of sub-records. Moreover, a number of additional hyperparameters need to be tuned, including the dimensions of the encoder network, as well as the length of historical data records used, increasing the burden on the user.

Alternatively, \emph{Unconstrained Multiple Shooting} (UMS), where initial states of sub-records are considered unconstrained optimization variables, is considered. Like EBMS, UMS does not require constrained optimizers, enables parallel evaluation of multiple sub-records and improves conditioning of the cost function. Unlike EBMS, UMS does not require tuning of additional hyperparameters or training with a large number of sub-records. In fact, as the number of learnable initial states scales linearly with the number of sub-records, complexity is lowest when few sub-records are considered. Conceptually, UMS and EBMS aim to solve ill-conditioning of cost functions through similar means, however the cost-benefit trade-off, in terms of complexity, is vastly different. In principle, EBMS favors cases with large availability of data, and UMS favors cases with less data and few sub-records, but scales poorly as the number of records increases. Some research on the reliability of these methods has been performed, \cite{forgione2022} analyzes variance of EBMS for different hyperparameters. Additionally, \cite{retzler2022} analyzes the effect of sub-record length for constrained methods. However, analysis on the variance of UMS is missing, and importantly, analysis on the aforementioned complexity trade-off between UMS and EBMS is missing.

To improve the reliability of jointly estimating LPV-LFR models and their scheduling maps, this work contributes:  effective learning of LPV models with Linear Fractional Representation through:\begin{enumerate}
    \item Analysis showing how rational-dependency degrades conditioning of the simulation cost.
    \item BLA initializations for LPV models with Linear Fractional Representation and well-posed rational parameterization.
    \item A numerical comparison of UMS and EBMS, highlighting their complexity trade-off.
\end{enumerate} 

The remainder of this paper is structured as follows: first, in Section \ref{sec: ident_setting} the problem setting and LPV-LFR model structure are introduced. Subsequently, in Section \ref{sec: conditioning} the conditioning of cost functions is analyzed for these model structures. Section \ref{sec:bla} introduces a BLA initialization procedure for LPV-LFR models. In Section \ref{sec:shooting-methods} two unconstrained shooting methods to improve the conditioning of cost functions in identification of LPV-LFR models are introduced. Next, Section \ref{sec: sim_example} details the numerical study used to analyze the effects of the proposed initialization and multiple shooting methods. The results of this study are shown in Section \ref{sec: results}, followed by further discussions in Section \ref{sec:discussion} after which final conclusions are drawn in Section \ref{sec: conclusion}.
\section{Problem Setting} \label{sec: ident_setting}
\subsection{Data-generating system}
We consider nonlinear \emph{discrete-time} (DT) data-generating systems with the following state-space representation:
\begin{equation}
\label{NLsys_eq}
\Sigma_\mathrm{o}: \left\{
\begin{aligned}
x_{k+1}  &= f(x_k,u_k),\\
y_k &= g(x_k,u_k) + e_k,
\end{aligned}
\right.
\end{equation}

where $k \in \mathbb{Z}$ is the discrete time, where $x_k \in  \mathbb{X}\subseteq \mathbb{R}^{n_\mathrm{x}}$ is the state of the plant,  $u_k \in \mathbb{U} \subseteq \mathbb{R}^{n_\mathrm{u}}$ is the input signal, $y_k \in \mathbb{Y} \subseteq \mathbb{R}^{n_\mathrm{y}}$ is the system output, with output-additive white noise  $e_k \in \mathbb{E} \subseteq \mathbb{R}^{n_\mathrm{y}}$. The input $u_k$ is assumed to be quasi-stationary. The state transition function $f: \mathbb{X} \times \mathbb{U} \!\to\! \mathbb{X}$ and output function $g: \mathbb{X} \times \mathbb{U} \!\to\! \mathbb{Y}$ are considered to be deterministic and nonlinear.

\subsection{LPV models with Linear Fractional Representation}
In this work we consider the class of LPV models with Linear Fractional Representations. Models follow well-posed-by-construction parameterizations as proposed in \cite{drenth2026}. Joint estimation of a \emph{Neural Network} (NN) scheduling map $\hat{\psi}_\theta$ and LPV-LFR model $G_\theta$ is performed. The LPV-LFR model has the following structure:
\begin{equation}
\label{LPVLFR_plant}
G_\theta: \left\{
\begin{aligned}
\begin{bmatrix}
\hat{x}_{k+1} \\
\hat{z}_k\\
\hat{y}_k
\end{bmatrix} &=
\begin{bmatrix}
A_x & B_w & B_u\\
C_z & D_{zw} & D_{zu}\\
C_y & D_{yw} & D_{yu}
\end{bmatrix}
\begin{bmatrix}
\hat{x}_k\\
\hat{w}_k\\
u_k
\end{bmatrix},\\
\hat{w}_k &= \Delta(\hat{p}_k)\hat{z}_k,
\end{aligned}
\right.
\end{equation}
where $\Delta(\hat{p}_k)$ is diagonal with repeating values of $\hat{p}_k,$ i.e.,  
\begin{equation}
\label{diag_delta}
\Delta(\hat{p}_k) = \mathrm{diag} ( \hat{p}_{k,1} I_{r_1}, \dots, \hat{p}_{k,n_p} I_{r_{n_p}}).
\end{equation}  $\hat{w}_k$ and $\hat{z}_k$ represent the latent variables of the LFR, eliminating these variables results in the equivalent  state-space representation with scheduling dependent matrices:
\begin{equation}\label{eq: collapsed_ss}
G_\theta: \left\{
\begin{aligned}
f(x_k, u_k, p_k) &= \mathcal{A}(p_k)x_k + \mathcal{B}(p_k)u_k,\\
g(x_k, u_k, p_k) &= \mathcal{C}(p_k)x_k + \mathcal{D}(p_k)u_k,
\end{aligned}\right.
\end{equation}
where $\mathcal{A}(p_k), \mathcal{B}(p_k), \mathcal{C}(p_k)$ and $\mathcal{D}(p_k)$ represent scheduling-dependent matrix functions. Models with rational dependency follow the well-posed parameterization of \cite{drenth2026}, i.e., $D_{zw} = e^{-\Psi S}$ with
\begin{equation*}
   \Psi = \mathrm{diag}(e^{D_d}), \quad S = D_A^\top D_A + D_B - D_B^\top + \epsilon_S I .
\end{equation*}
 A feedforward neural network with linear bypass, i.e., a ResNet, architecture is used for estimating the scheduling map $\hat{\psi}_\theta$.

The trainable parameters of the LPV-LFR model are collected in $\theta_G$ for the affine and rational cases:
\begin{align}
  \theta_G^{\mathrm{aff}} &= \mathrm{vec}\left(A_x, B_w, B_u, C_z, D_{zu}, C_y, D_{yw}\right), \label{eq:theta_affine}\\
  \theta_G^{\mathrm{rat}} &= \mathrm{vec}\left(\theta_G^{\mathrm{aff}}, D_A, D_B, D_d\right). \label{eq:theta_rational}
\end{align}

Finally, $\theta_\psi$ collects the trainable parameters of the scheduling map.

\section{Conditioning of LPV-LFR models}\label{sec: conditioning}
In \cite{ribeiro2020} an explicit relationship between the Lipschitz constant of the simulation error cost function, the Lipschitz constant of the state-transition function of a model and the length of the simulation is described. Specifically, the following is stated:
\begin{equation}\label{eq:lip_sim_cost}
    L_V = \begin{cases}
\mathcal{O}(L_f^{2T})   & \text{if } L_f > 1, \\
\mathcal{O}(T)          & \text{if } L_f = 1, \\
\mathcal{O}(1)          & \text{if } L_f < 1.
\end{cases}
\end{equation}
where $L_V$ represents the Lipschitz-constant of the simulation error cost function, $L_f$ represents the Lipschitz constant of the state-transition function with regard to $x_k$ and  its parameters $\theta$. The truncated simulation length is denoted as $T$. To motivate the use of multiple shooting methods for LPV-LFR models with rational scheduling-dependency, we analyze the Lipschitz constant of its state-transition function, described in \eqref{eq: collapsed_ss}. For simplicity, we illustrate this for the \emph{externally-scheduled} case, where the scheduling signal is considered measurable, but bounded. Nonetheless, the effect on conditioning is expected to hold generally, in fact, the dependency  of $p_k$ on $x_k$ in self-scheduled models is expected to worsen the results. Similarly, we do not consider the additional complexity of well-posed construction of $D_{zw}$.

\begin{assum}[Scheduling]
\label{ass:scheduling}
$\Delta(\hat{p}_k)$ has diagonal structure defined in \eqref{diag_delta} with $\|\hat{p}_k\|_\infty \leq 1$ hence $\|\Delta(\hat{p}_k)\| \leq 1$ for all $k$.
\end{assum}
\begin{assum}[Bounded trajectories]
\label{ass:bound_traj}
The state and input signals have bounded norms, i.e., $\|x_k\| \leq M_x, \quad \|u_k\| \leq M_u$ for all $k$.
\end{assum}

\begin{assum}[Parameter set]
\label{ass:param_set}
 $\theta_G\in\Theta_G$, where $\Theta_G\subset\mathbb {R}^{n_\theta}$ is compact
and convex, and there exist $\epsilon\in(0,1]$ and $\kappa_W\geq 1$ such that,
for every $\theta\in\Theta_G$, there is an invertible $W$ with
$W\Delta(p)=\Delta(p)W$ for all $p$, satisfying
\begin{equation*}
  \|W^{-1}D_{zw}W\|_2 \leq 1-\epsilon,\qquad \|W\|_2\|W^{-1}\|_2 \leq \kappa_W .
\end{equation*}   
\end{assum}

For conciseness, in the remainder, for any matrix $X$ of $\theta_G$, $\|X\|_2$ represents $\max_{\theta_G \in \Theta_G}\|X\|_2$ which is finite following Assumption \ref{ass:param_set}. 

\begin{lem}[Rational gain]
\label{lem:rat-gain}
Under Assumptions \ref{ass:scheduling} and \ref{ass:param_set}, $I-D_{zw}\Delta(p_k)$ is invertible for all
$p_k$ and $\theta\in\Theta_G$, and
$\Phi:=\Delta(p_k)\big(I-D_{zw}\Delta(p_k)\big)^{-1}$ satisfies
\begin{equation}
  \|\Phi(p_k,\theta)\|_2\le\gamma:=\kappa_W\,\epsilon^{-1}.
\end{equation}
\end{lem}

\begin{pf}
Let $\tilde D:=W^{-1}D_{zw}W$. Since $W$ commutes with $\Delta:=\Delta(p_k)$,
$I-D_{zw}\Delta=W(I-\tilde D\Delta)W^{-1}$. By Assumptions \ref{ass:scheduling} and \ref{ass:param_set},
$\|\tilde D\Delta\|_2\le1-\epsilon<1$, so the Neumann series gives
invertibility and $\|(I-\tilde D\Delta)^{-1}\|_2\le\epsilon^{-1}$. Hence
$\Phi=W\Delta(I-\tilde D\Delta)^{-1}W^{-1}$ and
$\|\Phi\|_2\le\|W\|_2\|W^{-1}\|_2\,\epsilon^{-1}\leq\gamma$.
\qed
\end{pf}

\begin{rem}
The well-posed parameterization $D_{zw}=e^{-\Psi S}$ satisfies Assumption \ref{ass:param_set}
with $W=\Psi^{1/2}$.
\end{rem}

\begin{prop}[LPV-LFR Lipschitz]
Let Assumptions \ref{ass:scheduling}-\ref{ass:param_set} hold and let $\gamma$ be as in Lemma \ref{lem:rat-gain}. Then, for all $p_k$
with $\|p_k\|_\infty\leq 1$, all $u_k$ with $\|u_k\|_2\le M_u$, all $x,x'$ with
$\|x\|_2,\|x'\|_2\le M_x$ and all $\theta,\theta'\in\Theta_G$,
\begin{equation*}
  \|f(x,u_k,p_k,\theta)-f(x',u_k,p_k,\theta')\|_2
  \le L_f\left\|\begin{bmatrix}x-x'\\\theta-\theta'\end{bmatrix}\right\|_2,
\end{equation*}
where $L_f:=L_x+\sum_iL_{\theta_i}$, with $L_x$ and $L_{\theta_i}$:
\begin{equation}\label{eq:lipschitz_total}
    \begin{aligned}
  L_x        &= \|A_x\|_2 + \gamma\, \|B_w\|_2 \|C_z\|_2, \\
  L_A        &= M_x, \quad  L_{B_u}    = M_u, \\
  L_{B_w}    &= \gamma\, \left(\|C_z\|_2 M_x + \|D_{zu}\|_2 M_u\right), \\
  L_{C_z}    &= \gamma\, \|B_w\|_2 M_x, \quad
  L_{D_{zu}} = \gamma\, \|B_w\|_2 M_u, \\
  L_{D_{zw}} &= \gamma^2\, \|B_w\|_2 \left(\|C_z\|_2 M_x + \|D_{zu}\|_2 M_u\right). &&
\end{aligned}
\end{equation}
\end{prop}

\begin{pf}
By Lemma \ref{lem:rat-gain}, $f=A_xx+B_uu_k+B_w\Phi\bar z$ with $\bar z:=C_zx+D_{zu}u_k$,
where $\|\Phi\|_2\leq\gamma$ and $\|\bar z\|_2\leq\|C_z\|_2M_x+\|D_{zu}\|_2M_u$.
For $x$,
\begin{align*}
  &\|(A_x+B_w\Phi C_z)(x-x')\|_2\\
  &\quad\leq\left(\|A_x\|_2+\gamma\|B_w\|_2\|C_z\|_2\right)\|x-x'\|_2\\
  &\quad=L_x\|x-x'\|_2 .
\end{align*}
For $\theta$, change one block $\theta_i$ at a time, with
$\delta\theta_i:=\theta_i-\theta_i'$ and $\Phi'$ the value of $\Phi$ at $\theta'$. By
sub-multiplicativity, and using $\Phi-\Phi'=\Phi\,\delta D_{zw}\,\Phi'$ for
$D_{zw}$, each change is bounded by
$L_{\theta_i}\|\delta\theta_i\|_2\le L_{\theta_i}\|\theta-\theta'\|_2$, with
$L_{\theta_i}$ as in (10). Summing gives $L_f$, and since $\|x-x'\|_2$ and
$\|\theta-\theta'\|_2$ are both bounded by the norm on the right-hand side
of \eqref{eq:lipschitz_total}, this proves \eqref{eq:lipschitz_total}. \qed
\end{pf}

In the affine-dependency case, $\gamma = 1$ and $L_{D_{zw}} = 0.$ The additional scaling of $L_f$ through $\gamma$ in \eqref{eq:lipschitz_total} clearly increases the bound of $L_v$ according to  \eqref{eq:lip_sim_cost} for models with rational dependency and small $\epsilon$. This stresses the need for methods that improve the conditioning of the cost function, such as multiple shooting.

\section{BLA-based initialization procedure}\label{sec:bla}

\subsection{BLA Initialization Procedure}\label{sec: bla_init}
In this work we adapt the initialization procedure of \cite{schoukens2020} for NL-LFR structures to the LPV-LFR case. Specifically, we extend the methods to support initialization of LPV-LFR models with well-posed rational scheduling dependency. First, a parametric BLA estimate with state-space representation must be obtained. In this work this is obtained by directly identifying an LTI state-space model which minimizes the mean-squared simulation error cost:
\begin{equation}\label{eq:SE-cost}
V(\theta) = \frac{1}{N}
  \sum_{k=0}^{N-1}\|y_k-\hat{y}_k\|_2^2,\\
\end{equation}
where $N$ is the simulation length and $\hat{y_k}$ represents the simulation output. However, many alternative approaches to obtain a BLA exist and can be used. For example, instead of directly estimating with PEM \citep{ljung1999}, other methods initialize using state-space subspace estimates \citep{schoukens2020}, or non-parametric and frequency-domain subspace methods \citep{paduart2010}.

 Having obtained the matrices of the BLA estimate\\ $(A_x^\mathrm{bla}\ , B_u^\mathrm{bla}\ , C^\mathrm{bla}_y\ ,D_{yu}^\mathrm{bla})$, the BLA is simulated on the training data, resulting in a simulated state trajectory $x(k)$. Subsequently, a similarity transform of the BLA is performed:
\begin{equation}\label{eq: bla_transform}
\begin{array}{lll}
\tilde{A_x}^\mathrm{bla} = T_xA_x^\mathrm{bla}T_x^{-1}, \qquad & \tilde{B}_u^\mathrm{bla} = T_xB_u^\mathrm{bla},\\
\tilde{C}_y^\mathrm{bla} = C_y^\mathrm{bla}T_x^{-1}, \qquad &\tilde{D}_{yu}^\mathrm{bla} = D_{yu}^\mathrm{bla} ,
\end{array}
\end{equation}
where the transformation matrix $T_x \in \mathbb{R}^{n_\mathrm{x} \times n_\mathrm{x}}$ is a diagonal matrix with the inverse sample standard deviation of each state, i.e., %
$T_x = \mathrm{diag}\left(\sigma(x_1(k))^{-1}, \ldots, \sigma(x_{n_\mathrm{x}}(k))^{-1}\right),$
with $\sigma(\cdot)$ representing the sample standard deviation. This transformation ensures the states of the transformed BLA model have unit variance, which is considered favorable as initial guess, as the state is an input of the NN scheduling map \citep{dehkordi2026}.

Secondly, based on the transformed system matrices of the BLA estimate, an LPV-LFR model is initialized using the following initialization strategy:
\begin{equation}
\begin{array}{ll}
\label{eq:LFR_init_BLA}
A_x = \tilde{A}_\mathrm{BLA}, \quad & B_w = \mathbf{0},\\
B_u = \tilde{B}_\mathrm{BLA}, \quad & C_z \sim \mathcal{U}(-0.1, 0.1),\\
D_{zu} \sim \mathcal{U}(-0.1, 0.1), \quad & C_y = \tilde{C}_\mathrm{BLA}, \\
D_{yw} = \mathbf{0}, \quad &D_{yu} = \tilde{D}_\mathrm{BLA}.
\end{array}
\end{equation}
Furthermore, for affine-dependency models, $D_{zw} = \mathbf{0}$. For rational-dependency models we additionally initialize $D_{zw} = 0.5I$, for a well-posed parameterization, this requires:
\begin{equation}\label{eq:lfr_init_dzw}
    D_A = \sqrt{-\ln(0.5)}  \cdot I, \qquad D_B = \mathbf{0}, \qquad D_d = \mathbf{0}.
\end{equation}
Ideally, initializing $D_{zw} = \mathbf{0}$ would be chosen, as this would correspond to affine-dependency and would reduce $\gamma$ in \eqref{eq:lipschitz_total} to one for the initialized model. However, the well-posed parameterization using the matrix exponential cannot represent $D_{zw} = \mathbf{0}$. Instead, the initialization relies on zero initialization of $B_w$ and $D_{yw}$ ensures latent channel does not affect the state and output of the initial model.

Finally, the scheduling map with ResNet structure is initialized using uniform Xavier initialization \citep{glorot2010} for the weight matrices, and zero initialization for all the bias vectors. Note that the Xavier initialization relies on inputs with unit variance, which is achieved by the state-transformation \eqref{eq: bla_transform} and IO normalization for the states and inputs respectively.

Note that the choice in initialization \eqref{eq:LFR_init_BLA} ensures the LPV-LFR model matches the behaviour of the BLA estimate at the first iteration. Consequently, the LPV-LFR is stable at initialization if the BLA estimate is stable. 

\section{Shooting methods} \label{sec:shooting-methods}

\subsection{Unconstrained Methods}
In this work, we consider UMS and EBMS. These methods are chosen as they are compatible with unconstrained optimization approaches, and only require forward and gradient calculations along each sub-record, unlike the constrained methods. Finally, UMS and EBMS also support multiple shooting with overlapping sub-records, potentially enjoying benefits in terms of \emph{data-efficiency} \citep{beintema2023}.

\subsection{Truncated Prediction Error}
Incorporating EBMS and UMS into the identification process requires substituting the prediction error \eqref{eq:SE-cost}  with the truncated prediction error cost:
\begin{subequations}\label{eq:MS}
\begin{align}
\min_{\theta}\;\; & V^\mathrm{MS}(\theta) = \frac{1}{T N_\mathrm{rec}}
  \sum_{t\in\mathcal I}\sum_{k=0}^{T-1}\big\|y_{t+k}-\hat y_{t+k|t}\big\|_2^2 \label{eq:MS-cost}\\
\text{s.t.}\;\;
& \hat x_{t+k+1|t} = f_\theta\left(\hat x_{t+k|t},\,u_{t+k}\right), \label{eq:MS-state}\\
& \hat y_{t+k|t} = g_\theta\left(\hat x_{t+k|t},\,u_{t+k}\right), \label{eq:MS-output}\\
& \hat x_{t|t} = x_{0,t}, \qquad \forall\, t\in\mathcal I, \label{eq:MS-init}
\end{align}
\end{subequations}
where $T$ denotes the truncation length, i.e., the sample-length of each sub-record and $\mathcal{I}$ is the set of indices corresponding to each sub-record. $N_\mathrm{rec}$ denotes the number of sub-records, equal to the cardinality of $\mathcal{I}$, i.e.,  $\mathrm{card}(\mathcal{I}) = N_\mathrm{rec}$. The initial states are generated depending on the type of multiple shooting used:
\begin{equation}\label{eq:x0}
x_{0,t} =
\begin{cases}
\phi_\theta\left(y_{t-n_a:t-1},\,u_{t-n_b:t-1}\right), & \text{EBMS}\\
\theta_{x_0,t}, & \text{UMS}
\end{cases}
\end{equation}
where $\theta_{x_0,t}$ collects the trained initial states and $\phi_\theta$ represents the encoder network, mapping the past $n_a$ outputs and $n_b$ inputs to the initial state of each sub-record. In this work, encoder networks with ResNet structure are considered, its parameters collected in $\theta_\phi$. Observe that the single-shooting simulation error \eqref{eq:SE-cost} is a special case of UMS. Indices have equidistant spacing and are generated as follows:
\begin{equation}
    \mathcal{I} = \left\{n_\mathrm{hist} + j\,(T - n_\mathrm{o}) \;\middle|\; j = 0, 1, \dots, N_\mathrm{rec} - 1 \right\},
\end{equation}
where $n_\mathrm{o}$ denotes the number of overlapping samples and $n_\mathrm{hist} = \max\{n_a, n_b\}$ for EBMS and $n_\mathrm{hist} = 0 $ for UMS.
\section{Numerical experiments} \label{sec: sim_example}
\begin{figure*}[t]
  \centering
  \includegraphics[width=\textwidth]{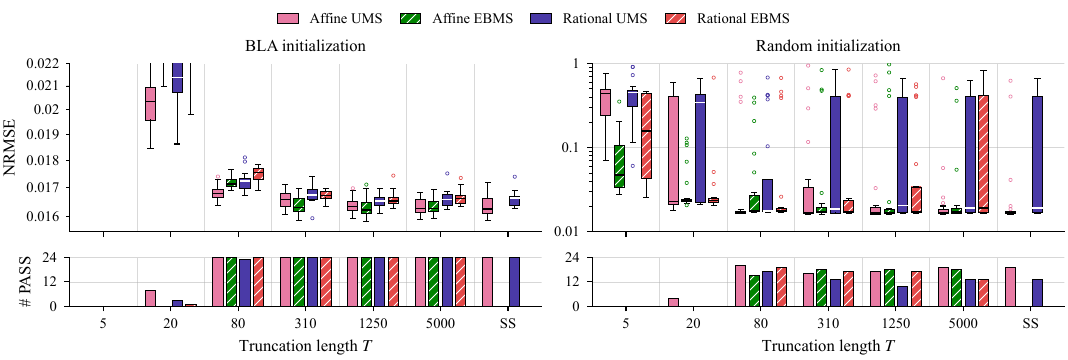}
  \caption{Accuracy of trained models versus length of sub-records, evaluated on test dataset. UMS configurations without overlap are shown, EBMS configurations have maximum overlap $(93.75\%)$.}
  \label{fig:boxplot_comp}
\end{figure*}

\subsection{Nonlinear Mass-Spring-Damper}
To compare the effect of initialization and shooting approaches, we consider a numerical example\footnote{A link to the git repository will be added before final revision.} based on the discrete-time \emph{NonLinear-Mass-Spring-Damper} (NL-MSD) introduced in \cite{drenth2026}. The dynamics of the NL-MSD are described by the following equations:
\begin{align}
\label{eq: DT_NL_MSD_eq}
x_{1,k+1} \!&=\! x_{1,k} \!+\! T_sx_{2,k},\notag\\
x_{2,k+1} \!&=\! x_{2,k} \!+\! \tfrac{T_s}{m}\left(u_k \!-\! \tfrac{3}{5}x_{1,k}u_k \!-\! k_1 x_{1,k} \!-\! k_2x_{1,k}^3 \!-\! d_1 x_{2,k} \right),\notag\\
y_k \!&=\! x_{1,k},
\end{align}
with $T_s \! =\! 0.1$, $m \!= \! 1$, $k_1 \!=\! 0.1$, $k_2 \!=\! 1$, $d_1 \! = \! 1$ and where $x_1$ represents the position of the system and $x_2$ its velocity. The nonlinearities in \eqref{eq: DT_NL_MSD_eq} are dependent on $x_{1,k}u_k$ and $x_{1,k}^3$, requiring two scheduling variables, $p_{1,k} = x_{1,k}, \; p_{2,k} = x_{1,k}^2$, for affine LPV representations and one scheduling variable, $p_{1,k} = x_{1,k}$, for rational dependency models.

Three datasets are generated using the NL-MSD, one training, validation and test set with lengths $N_\mathrm{train} = N_\mathrm{test} = 2 \cdot 10^4$ and $N_\mathrm{val} = 5 \cdot 10^3$. For each dataset, the system is excited with a different realization of zero-mean random phase multisine exciting all frequencies from 0 to $1 \; \mathrm{Hz}$, $0 \; \mathrm{Hz}$ excluded. The period length of the multisine is equal to the total length of the training, validation and test set respectively. The train and validation inputs have an RMS amplitude of 1. The input of the test data set has RMS amplitude of 0.9, ensuring the test data does not go out of distribution. The generated outputs of the train and validation are artificially contaminated with additive Gaussian white noise, resulting in train and validation datasets with SNR of 40 dB. The test dataset is kept noise-free for analysis.

\subsection{Configurations}
We analyze performance of two model configurations: rational and affine dependency LPV-LFR models. As shown in \cite{drenth2026}, affine models with ($n_\mathrm{x} \!=\! 2, n_\mathrm{p} \!=\! 2, r_1 = r_2 \!=\! 1$) and rational models with ($n_\mathrm{x} \!=\! 2, n_\mathrm{p} \!=\! 1, r_1 \!=\! 3$), are capable of accurately modelling the NL-MSD, these same model orders are used in this work.
Moreover, we consider linear scheduling maps with saturation:
\begin{equation}
\label{eq: MSD_schedmap}
    \hat{\psi}(x,u) = \tanh(W_\mathrm{x}x + W_\mathrm{u}u + b).
\end{equation}
Besides model type, we consider two initialization methods. The BLA initialization strategy proposed in Section \ref{sec: bla_init} is compared with a random initialization strategy. Random initialization follows the same procedure as \eqref{eq:LFR_init_BLA} and \eqref{eq:lfr_init_dzw}, but substitutes the BLA estimate for the following random initialization strategy for $A_x, B_u, C_y$ and $D_{yu}$:
\begin{equation}
\begin{array}{ll}
\label{LFR_init_BLA}
A_x = 0.5I, \quad & B_u \sim \mathcal{N}(0,1)\\
C_y \sim \mathcal{N}(0,1), \quad & D_{yu} = \mathbf{0}.\\
\end{array}
\end{equation}
Moreover, two different multiple shooting methods are considered: UMS and EBMS. Experiments are performed across a range of sub-record lengths, $T \in \{20000, 5000, \allowbreak 1250, 310, 80, 20, 5\}$. Furthermore, four overlap percentages are considered, $n_\mathrm{o}(\%) \in \{0, 50, 75, 93.75\}$. EBMS requires training an encoder network, all encoder networks considered are parameterized as 2-hidden layer network with $\mathrm{tanh}$ activation functions, 12 nodes per layer and a linear bypass. History lengths $n_a =n_b = 4n_x +1$ are used. The parameters of the encoder networks are initialized using Xavier initialization.
Combining the model types, initialization methods, shooting methods, record lengths and overlap percentages results in a total of 188 unique configurations. To draw meaningful conclusions, 24 models are trained per configuration, each model initialized from a different seed.

\subsection{Model construction}
All models are trained following the same optimization procedure, according to the truncated prediction error \eqref{eq:MS}. Optimization of this cost follows the two-step approach proposed in \cite{bemporad2025a}. A 1000 iterations of Adam provide a warm start for the \emph{limited-memory Broyden-Fletcher-Goldfarb-Shanno} (L-BFGS) algorithm of at most 10 000 iterations. Furthermore, \emph{early-stopping} is used as regularization. In this method, the model with the lowest simulation error on the validation data is selected during training, preventing overfitting on the training data. Models and optimization are implemented in Python, leveraging the computational efficiency of JAX.
\section{Results}\label{sec: results}
Evaluation of the results is performed based on the \emph{Normalized Root-Mean-Square Error} (NRMSE) defined as:
\begin{equation}
\mathrm{NRMSE} = \sqrt{\frac{\sum_{k=1}^{N}\left(y_k-\hat{y}_k\right)^2}{\sum_{k=1}^{N}\left(y_k-\bar{y}\right)^2}},
\end{equation}
where $\bar{y}$ is the sample mean of $y$.

First, Figure \ref{fig:boxplot_comp} shows the following:
(i) both model types and initialization methods are capable of reaching a low NRMSE ( $\leq 0.02$); (ii) the initialization strategy has the largest influence on achieved results and reduces the spread of results in the rational case. Possibly, the BLA initialization starts optimization in a well-conditioned part of the loss landscape, improving subsequent refinement; (iii) no clear differentiation between the spread of results of rational and affine models is observed in the BLA initialization case; (iv) randomly initialized rational models benefit in a reduced spread of results as multiple shooting is used. In particular, EBMS strongly reduces the spread, outperforming UMS for nearly each $T$; (v) sub-record lengths of 5 and 20 are too short to accurately capture the dynamics of the data-generating system.

\begin{figure}[t]
\begin{center}
  \includegraphics[width=\columnwidth]{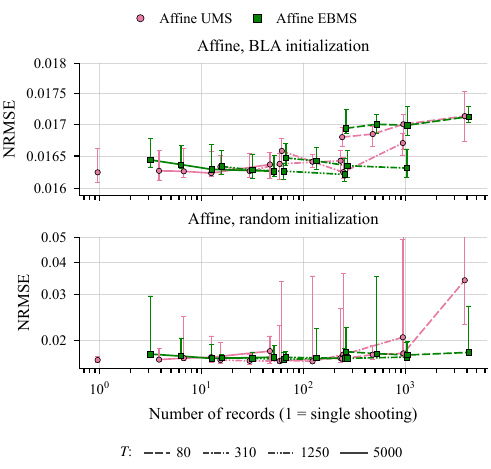}
  \caption{Accuracy of trained affine models versus number of sub-records, evaluated on test dataset. Results show median and IQR of BLA initialized models and randomly initialized models. $T \in \{5, 20\}$ are omitted.}
  \label{fig:aff_iqr_nrecords}
\end{center}
\end{figure}

In the next figure, accuracy as function of the number of records is examined. In Figure \ref{fig:aff_iqr_nrecords} two things are apparent: (i) using BLA initialization, the number of records does not meaningfully change the results. Instead, the only outliers in results centered around NRMSE of $1.7\%$ are attributed to truncation length $T = 80$; (ii) random initializations show that the spread of obtained results increases more quickly for UMS than for EBMS, as the number of records increases. For $T=80$ this even affects the attainable results, as the models trained with UMS lose accuracy more strongly (in the median) than EBMS models.

\begin{figure}[t]
\begin{center}
  \includegraphics[width=\columnwidth]{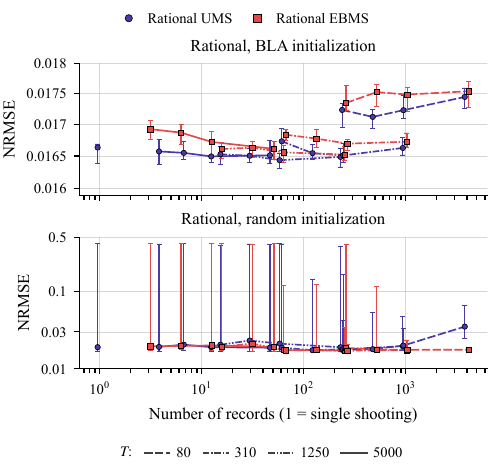}
  \caption{Accuracy of trained rational models versus number of sub-records, evaluated on test dataset. Results show median and IQR of BLA initialized models and randomly initialized models. $T \in \{5, 20\}$ are omitted.}
  \label{fig:rat_iqr_nrecords}
\end{center}
\end{figure}

Similar trends are shown in the rational case in Figure \ref{fig:rat_iqr_nrecords}, the behaviour of the BLA initialized methods follows the pattern of the affine case. However, the randomly initialized models show a more interesting trend. Both MS methods show reduced spread as the number of records increases, attributable to the improved conditioning of the cost for shorter sub-records, but the spread of EBMS methods keeps decreasing, as the median and IQR of the UMS method worsens. We conjecture that, in the case with the largest number of sub-records of EBMS, two effects occur: (i) the shorter sub-records improve conditioning of the cost function; (ii) the large amount of overlap increases the variance of the estimates in the UMS case, but decreases the variance of the estimates in the EBMS case, as the number of records scales with the overlap, improving the training of the encoder.

\begin{figure}[b]
\begin{center}
  \includegraphics[width=\columnwidth]{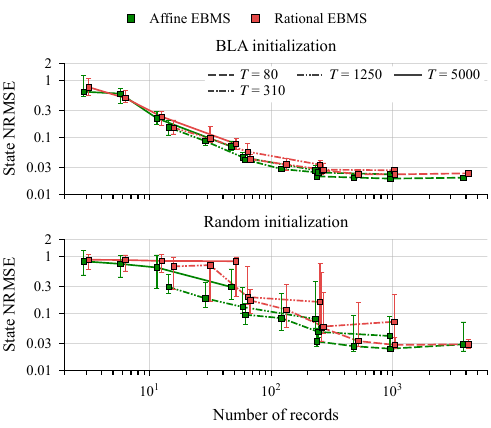}
  \caption{State-estimation accuracy of encoders versus number of records, evaluated on test dataset. Results show median and IQR for BLA initialized models (top) and randomly initialized models (bottom). $T \in \{5, 20\}$ are omitted.}
  \label{fig:encoder_state_comp}
\end{center}
\end{figure}

Next, we evaluate how accurately the encoder is able to capture the state-trajectory. We evaluate the averaged NRMSE of the simulated states of a model, and the states estimated by the encoder when used as a dead-beat state-estimator. A comparison against number of sub-records is shown in Figure \ref{fig:encoder_state_comp}. BLA initialized and randomly initialized models follow the same trend. The state NRMSE and its IQR decrease as the number of records increases, matching expected behaviour. Interestingly, the actual NRMSE of the predictions in figures \ref{fig:aff_iqr_nrecords} and \ref{fig:rat_iqr_nrecords} does not show this same trend indicating that for lower number of sub-records, the encoder network is able to learn initial state estimates of the training set or that estimating initial states accurately has little importance for long sub-records on the considered example.

\begin{figure}[ht]
\begin{center}
  \includegraphics[width=\columnwidth]{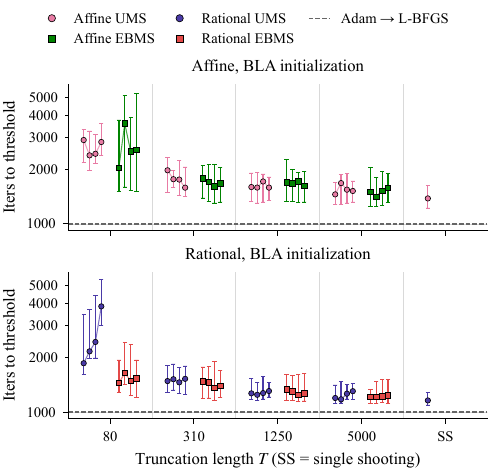}
  \caption{Number of iterations required to reach validation NRMSE threshold of $2.2 \;\%$. $T \in \{5, 20\}$ are omitted.}
  \label{fig:convergence_results}
\end{center}
\end{figure}

Finally, we analyze the impact of record length and overlap on convergence rate. To this end, we set a threshold on the validation NRMSE during training at $2.2\%$ to which nearly all models in the BLA case converge. Next, we display the number of iterations required to reach this threshold. The results are shown in Figure \ref{fig:convergence_results}. We make two observations: (i) despite the theoretical increase in data-efficiency, increasing the overlap has no meaningful impact on convergence rate; (ii) when using a BLA initialization, MS almost always has negative effect, or no positive impact, on the convergence rate of the considered example. However, it must be noted that this is convergence rate in terms of iterations, in terms of wall-clock time, MS provides theoretical benefit over single-shooting, given parallel compute can be leveraged.
\section{Discussion} \label{sec:discussion}
Selecting the best model out of a sufficient number of random initializations can work well to achieve accurate estimates of nonlinear systems. However, as model complexity grows, this might become computationally challenging. Instead, our results show that initializing LPV models with a BLA estimate can result in sufficiently accurate results. Instead of training multiple models, parallel compute is then freed to instead reduce training time, made possible by multiple shooting.

When a sufficiently accurate BLA is not available, UMS and EBMS are capable of reducing the spread of achieved results, reducing need for multiple initializations. Although only one simulation example is considered in this work, by analyzing two model structures with different conditioning of the cost function, important insights can still be gained. In particular we note that in general UMS provides a simple to tune, computationally efficient method for a relatively small number of sub-records. In cases where short sub-record are required to improve conditioning of the cost function, such as for rational LPV-LFR models, EBMS methods provide a solution that does not sacrifice achievable performance or spread, but comes at the cost of more user burden and additional complexity. 

Finally, it should be noted that EBMS methods are typically used in conjunction with mini-batching. Due to incompatibility with traditional L-BFGS methods, this was not considered in this work, however extensions of L-BFGS that support mini-batching exist \citep{berahas2016} and could improve effectiveness of EBMS beyond what is shown in this work. It is however expected that UMS would not benefit much from mini-batching, as each initial state needs to be updated individually.


\section{Conclusion} \label{sec: conclusion}
This work has analyzed the conditioning of cost landscapes of LPV-LFR models. A BLA initialization method which reduces the impact of the conditioning has been proposed and analyzed through numerical experiments, showing major improvement over random initialization in the considered example. Furthermore, two unconstrained multiple shooting methods have been compared in a numerical study, both showing improvements in learning of randomly initialized models. Furthermore, the complexity trade-off between these two methods has been illustrated, leading to recommendations for the user. 

\bibliography{ifacconf} 
\appendix

\end{document}